\documentstyle[aps,amsmath,psfig,floats]{revtex}
\newcommand{\eV}{{\rm eV}}
\newcommand{\MeV}{{\rm MeV}}

\begin{document}

\title{Tritium Decay and the Hypothesis of Tachyonic Neutrinos}

\author{Jacek Ciborowski}
\address{Department of Physics, University of Warsaw, ul.\ Ho\.{z}a
  69, PL-00-681 Warsaw, Poland}
\author{Jakub Rembieli\'nski}
\address{Department of Theoretical Physics, University of \L\'od\'z, ul.\ 
  Pomorska 149/153, PL-90-236 \L\'od\'z, Poland}

\wideabs{\maketitle
\begin{abstract}Numerous recent measurements  indicate an
excess of counts near the endpoint of the electron energy spectrum
in tritium decay. We show that this effect is expected 
if the neutrino is a tachyon. Results of  calculations, 
based on a unitary (causal)  theory of tachyons, are  presented.
The hypothesis of tachyonic neutrinos also offers a natural explanation
of the V--A structure of the weak leptonic current
in neutrino interactions.
\end{abstract}}


\narrowtext
\section{Introduction}\label{section:introduction}

Since several years numerous experiments have been performed with the
aim of measuring the electron antineutrino mass in tritium decay,
$^{3}{\rm H} \rightarrow \,^{3}{\rm He} + e^{-} +
\bar{\nu}_{e}$ \cite{mainz,troitsk,stoeffl95,other}.  This quantity
squared, $\xi = m^{2}_{\bar{\nu}_{e}}$, may be determined by
fitting the electron energy spectrum near the end point with the
formula given below in a simplified form:
\begin{equation}
  \frac{d\Gamma}{dE} \sim p (E_{\rm max} - E)\sqrt{(E_{\rm max} -
    E)^{2} - \xi},
\label{eq:old}
\end{equation}
where $E$ ($E_{\rm max}$) denote (maximal) energy of the electron in
this decay and $p$ its momentum.  Surprisingly, all recent experiments
yielded negative values of the parameter $\xi$.  Owing to the
increasing resolution of modern spectrometers, the reason of such
results has been attributed to a peculiar unexpected shape of the
spectrum in the vicinity of the end point. In qualitative term this
phenomenon, hereafter referred to as the {\it end point effect}, can
be viewed as an excess of counts in that region. This is contrary to
common expectations since if the neutrino were massive ($\xi>0$) a
depletion of counts towards the end point would be expected as
compared to the massless neutrino case.  The enhancement under
consideration has been found in the spectra collected at
Mainz~\cite{mainz}, Troitsk~\cite{troitsk} and earlier at
LLNL~\cite{stoeffl95}.  In particular, numerous measurements performed
by the two former groups deliver firm evidence in favour of the effect
which may be considered as well established experimentally.  Attempts
to determine the electron neutrino mass using formula~(\ref{eq:old}),
without additional assumptions concerning the origin of the
enhancement, lead to doubtful results in these circumstances.

The {\it end point effect\/} has not been convincingly explained on
conventional grounds.  Both Mainz and Troitsk groups made significant
efforts towards understanding their apparatus, data evaluation methods
and considered a wide class of related physical phenomena.  Moreover,
dedicated studies have demonstrated that the {\it end point effect\/}
could not originate from mistreatment of molecular effects~\cite{fsi}.
A possibility that it might be related to methods of evaluating the
data near the end of the physical region has also been
considered~\cite{stat}.  Lack of a credible explanation of the effect
made room for unconventional hypotheses~\cite{unconv}.

In this paper we present calculations of the electron energy spectrum
in tritium decay assuming that the neutrino is a tachyon. It must be
stressed that changing the sign in front of the parameter $\xi$
in~(\ref{eq:old}) does not at all convert it to the correct formula
describing a beta decay spectrum with a tachyonic neutrino.

A tachyon is a particle which moves with velocities always greater
than $c$, relative to any reference frame. The energy-momentum
relation reads: $E^{2} - \vec{p}^{2} = -\kappa^{2}$, where
$\kappa$ will be called the {\it tachyonic mass}.  Tachyons cannot be
described within the framework of the Einstein--Poincar\'e (EP)
relativity because of causality violation (this has been a repeated
argument to reject them as possibly existing particles).  It proved
also impossible to construct a unitary field theory of tachyons on
these grounds.  The unitary (causal) theory of tachyons proposed
recently \cite{jaremb}, free of these difficulties, is the basis for
calculations presented in this paper.  This theory does not invalidate
nor modify the EP theory of relativity for massive and light-like
particles.  In what follows the term `neutrino' stands for `electron
neutrino' or `electron antineutrino'. We use the following symbols:
total particle energy (momentum) -- $E(p)$; kinetic energy -- $T$; end
point energy -- $E_{\rm max}$, $T_{\rm max}$, respectively.

\section{Theory of spin--$\frac{1}{2}$ tachyons }\label{section:theory}

Since the time synchronisation scheme is a convention in the EP
relativity, therefore there is a freedom in the definition of the
coordinate time.  The standard choice is the Einstein--Poincar\'e
synchronisation with the one-way velocity of light isotropic and
constant.  This choice leads to the well known form of the Lorentz
group transformations but the EP coordinate time implies covariant
causality for time-like and light-like trajectories only. In order to
describe tachyons a different synchronisation scheme must be chosen,
namely that of Chang--Tangherlini (CT), preserving invariance of the
notion of the instant-time hyperplane \cite{ct}.  In this
synchronisation scheme the notion of causality is universal, i.e.
space-like trajectories (tachyons) are physically admissible too, the
only inconvenience being that the Lorentz group transformations, which
incorporate transformation rules for the velocity of a distinguished
(preferred) reference frame, have a more complicated form.  {\em The
  EP and CT descriptions are equivalent for the time-like and
  light-like trajectories; however a consistent (causal) description
  of tachyons is possible only in the CT scheme}.  A very important
consequence is that if tachyons exist then the relativity principle is
broken, i.e. there exists a preferred frame of reference, however the
Lorentz symmetry is preserved.  The interrelation between EP
($x_{\mathrm{EP}}$) and CT ($x$) coordinates reads:
\begin{equation}
  x_{\mathrm{EP}}^{0}=x^0+u^0 \vec{u} \vec{x},\quad
  \vec{x}_{\mathrm{EP}}=\vec{x},
\label{eq:theor1}
\end{equation}
where $u^{\mu}$ is the four-velocity of the privileged frame as seen
from the frame ($x^{\mu}$).  On the basis of these considerations a
fully consistent, Poincar\'e covariant quantum field theory of
tachyons has also been proposed~\cite{jaremb}.  In the case of the
fermionic tachyon with helicity $\frac{1}{2}$ the corresponding free
field equation reads:
\begin{equation}
  \left(\gamma^5\left(i\gamma\partial\right)-\kappa\right)\psi=0,
\label{eq:theor2}
\end{equation}
where the bispinor field $\psi$ is simultaneously an eigenvector of
the helicity operator with the eigenvalue $\frac{1}{2}$.  The elementary
tachyonic states are thus labelled by helicity.  The $\gamma$-matrices
are expressed by the standard ones in analogy to the relations
(\ref{eq:theor1}). The solution of the equation (\ref{eq:theor2}) is
given by:
\begin{multline}
  \psi(x, u) = \\
  \frac{1}{(2\pi)^{\frac{3}{2}}} \int d^4k\,\delta(k^2+\kappa^2)
  \theta(k^0)\left[w(k,u) e^{ikx}b^\dagger(k)\right.\\
  \left. +v(k,u) e^{-ikx}a(k)\right], \label{eq:theor3}
\end{multline}
where the operators $a$ and $b$ correspond to neutrino and
antineutrino, respectively. The amplitudes $v$ and $w$ satisfy the
following conditions:
\begin{gather}\label{eq:w-w}
  w(k, u) \bar{w}(k, u) = (\kappa-\gamma^5k\gamma)
  \frac{1}{2}\left(1-\frac{\gamma^5 [k \gamma, u \gamma]}{2
      \sqrt{q^2+\kappa^2}}\right)\\
  \label{eq:v-v}
  v(k, u) \bar{v}(k, u) = -(\kappa+\gamma^5k\gamma)
  \frac{1}{2}\left(1-\frac{\gamma^5 [k \gamma, u \gamma]}{2
      \sqrt{q^2+\kappa^2}}\right)\\
  \label{eq:n1}
  \bar w(k,u)\gamma^5 u \gamma w(k,u) =\bar v(k,u)\gamma^5 u \gamma
  v(k,u)=2q
  \\
  \label{eq:n2}
  \bar w(k^{\varPi},u)\gamma^5 u \gamma v(k,u)=0.
\end{gather}
Here $q=u_{\mu}k^{\mu}$ is equal to the energy of the tachyon in the
preferred frame and $\varPi$ denotes the space inversion operation.
In the massless limit, $\kappa\rightarrow 0$, the above relations are
identical with those obtained in the Weyl's theory.  An equation
similar in its form to (\ref{eq:theor2}) has already been proposed in
the EP synchronization~\cite{chodos85} however the theory based on the
latter is not unitary.

\section{Beta decay with a tachyonic electron neutrino}
\label{section:spectrum}

\subsection{Amplitude}

On the grounds of the formalism presented in sec.~\ref{section:theory}
we calculate the amplitude for a $\beta$ decay, $n\rightarrow p^{+} +
e^{-} + \bar{\nu}_{e}$, with a tachyonic electron antineutrino, using
an effective four--fermion interaction.  In the rest frame of the
decaying particle the decay rate for this process reads:
\begin{equation}
  d\Gamma=\frac{1}{4 m_{n} (2\pi)^5} d\Phi_3 \left|M\right|^2,
\label{eq:beta1}
\end{equation}
where $d\Phi_3$ is the phase-space volume element:
\begin{multline}
  d\Phi_3=\theta(k^0)\,\theta(l^0)\,\theta(r^0)
  \delta(k^2-m_{p}^{2})\,\delta(l^2-m_{e}^{2})\times\\
  \delta(r^2+\kappa^2)\, \delta^4(p-k-l-r)\, d^4k\, d^4l\, d^4r.
  \label{eq:beta2}
\end{multline}
The amplitude squared, $|M|^2$, can be derived (on the tree level)
directly from the lepton-hadron part of the effective Fermi
weak-interaction Lagrangian:
\begin{equation}
  \mathcal{L}_I = -G_F j_\mu J^\mu,
\end{equation}
where $j_\mu$ and $J^\mu$ denote leptonic and hadronic currents,
respectively.  However, under the condition that in the limit of the
zero neutrino mass the leptonic current takes the standard V--A form,
we have two natural choices, which we denote $helicity$ and
$chirality$ coupling.  Namely, we can choose the corresponding part of
the leptonic current in the form:
\begin{align}\label{hel:coupl}
  &\bar{u}_e \gamma^\mu w& \quad &\text{(helicity coupling)}
  \intertext{or} 
  &\bar{u}_e \gamma^\mu {\textstyle\frac12} (1 -
  \gamma^5) w & &\text{(chirality coupling)},
\end{align}
respectively.  In the former case the square of the amplitude ($M
\equiv M_{\rm h}$) reads:
\begin{multline}
  \left|M_{\rm h}\right|^2=2 \,G_{F}^{2}\, {\rm Tr}\left[u_e
    \bar{u}_e \gamma^{\mu} w \bar{w}
    \gamma^{\nu}\right] \times\\
  {\rm Tr}\left[u_p \bar{u}_{p}
    \gamma_{\mu}\left(1-g_A\gamma^5\right) u_n \bar{u}_n
    \gamma_{\nu}\left(1-g_A\gamma^5\right)\right],
\label{eq:beta3}
\end{multline}
while in the latter case (chirality coupling: $M \equiv M_{\rm ch}$):
\begin{multline}
  \left|M_{\rm ch}\right|^2=2 \,G_{F}^{2}\, {\rm Tr}\left[u_e
    \bar{u}_e \gamma^{\mu} {\textstyle\frac12} (1 - \gamma^5) w
    \bar{w}
    \gamma^{\nu} {\textstyle\frac12} (1 - \gamma^5) \right] \times\\
  {\rm Tr}\left[u_p \bar{u}_{p} \gamma_{\mu}\left(1-g_A\gamma^5\right)
    u_n \bar{u}_n \gamma_{\nu}\left(1-g_A\gamma^5\right)\right].
\label{eq:beta3a}
\end{multline}
Here $p$, $k$, $l$, $r$ are the four-momenta of $n$, $p^+$, $e^-$ and
$\bar{\nu}$ respectively; the corresponding masses are denoted by
$m_n$, $m_p$, $m_e$ and $\kappa$.
$G_F$ and $g_A$ are the Fermi constant and the axial coupling
constant. The amplitudes $u_n$, $\bar{u}_n$, $u_p$, $\bar{u}_p$,
$u_e$, $\bar{u}_e$ satisfy usual\footnote{But with $\gamma$ in the CT
  synchronisation \protect{\cite{jaremb}}.} polarisation relations:
$u_n \bar{u}_n=p\gamma+m_n$, $u_p \bar{u}_p=k\gamma+m_p$, $u_e
\bar{u}_e=l\gamma+m_e$, whereas $w \bar{w}$ is given by
Eq.~(\ref{eq:w-w}).  After elementary calculations
Eqs.~(\ref{eq:beta3}), (\ref{eq:beta3a}) read:
\begin{equation}
  \label{eq:beta4}
  \begin{split}
    |M_{\rm h}|^2 &= 16 G_F^2 \times \\
    & \times \Biggl\{\left[m_n m_p (1 - g_A^2) - k p (1 +
      g_A^2)\right] \Bigl(4 m_e \kappa \\
    &\quad - \frac{1}{\sqrt{(u r)^2 +
        \kappa^2}}\left[4 (\kappa^2 l u + l r \cdot u r) \right.\\
    &\qquad \left.- 2 \kappa^2 l u - 2 u r \cdot l r\right]\Bigr)\\
    &+ (1 + g_A^2) \Bigl(2 m_e \kappa (p k) \\
    &\quad - \frac{1}{\sqrt{(u r)^2 + \kappa^2}}\left[2 p k (k^2 l u +
      l r
      \cdot u r)\right.\\
    &\qquad \left. - 2 \kappa^2 (p l \cdot u k + k l \cdot u p ) - 2 u
      r (p l
      \cdot k r + k l \cdot p r)\right]\Bigr) \\
    &+ 4 g_A \Bigl(p r \cdot k l - p l \cdot k r \\
    &\quad + \frac{m_e \kappa}{\sqrt{(u r)^2 + \kappa^2}} (k r \cdot u
    p - p r \cdot u k)\Bigr)\Biggr\}
  \end{split}
\end{equation}
and
\begin{equation}
  \label{eq:beta4a}
  \begin{split}
    |M_{\rm ch}|^2 &= 16 G_F^2 \times \\
    &\times \Biggl\{\left(1 + \frac{u r}{\sqrt{(u r)^2 +
          \kappa^2}}\right) \Bigl[(g_A^2 - 1) m_n m_p l r \\
    &\quad + (g_A^2 + 1) (l p \cdot k r + p r \cdot k l)\\
    &\quad + 2 g_A (k l \cdot p r - l p \cdot k r)\Bigr]\\
    &+ \frac{\kappa^2}{\sqrt{(u r)^2 + \kappa^2}} \Bigl[(g_A^2 - 1)
    m_n m_p u l \\
    &\quad  + (g_A^2 + 1) (l p \cdot u k + u p \cdot k l)\\
    &\quad + 2 g_A (k l \cdot u p - l p \cdot u k)\Bigr]\Biggr\}
  \end{split}
\end{equation}

In order to calculate the differential energy spectrum of electrons in
the $\beta$ decay with a tachyonic electron neutrino, $d\Gamma/d l^0$,
it is necessary to account for the velocity of the preferred frame,
$u$.  We take $u_\mu = (1, 0, 0, 0)$ for simplicity, i.e. we derive
the result in a reference frame which is at rest with respect to the
preferred frame (consequences of a non-negligible velocity of the
preferred frame are discussed in Sec.~\ref{sec:Time_dep}).  The
spectrum $d\Gamma /d l^0$ may be obtained for both considered cases
(helicity and chirality coupling) by means of formulae
(\ref{eq:beta1}), (\ref{eq:beta2}), (\ref{eq:beta4}) and
(\ref{eq:beta4a}), after elementary integration with respect to
$d^4k\, d^3\vec{l}\, d^3\vec{r}$, from the following formula (which
gives identical results in the limit $\kappa^{2}\rightarrow 0$ as that
for the massless neutrino):
\begin{equation}\label{eq:dGdE1}
  \frac{d\Gamma}{d l^0}= \frac{1}{128 \pi^3 m_n}
  \int^{r_{+}}_{\max{(r_{-},0)}} d r^0\, \left|M(l^0,r^0)\right|^2
\end{equation}
with
\begin{multline*}
  r_{\pm}=\Bigl\{-\Delta m^2 l^0 + \Delta m^2 m_n + 2 (l^0)^2 m_n - 2
    l^0 m_{n}^{2}\\
  \pm \sqrt{(l^0)^2 - m_{e}^{2}} \left[(\Delta m^2)^2 + 4
      \kappa^2 m_{e}^{2} - 4 \Delta m^2 l^0 m_n\right. \\
   \left. - 8 \kappa^2 l^0 m_n + 4 \kappa^2 m_{n}^{2} + 4 (l^0)^2
      m_{n}^{2}\right]^{\frac12}\Bigr\} \times \\
  \times \left[2(m_{e}^{2} - 2 l^0 m_n + m_{n}^{2})\right]^{-1}.
\end{multline*}
Here $|M|^2 = |M_{\rm h}|^2$ or $|M_{\rm ch}|^2$, respectively, and $u
l = l^0$, $u r = r^0$, $u p = m_n$, $u k = m_n - l^0 - r^0$, $k p =
m_n (m_n - l^0 - r^0)$, $k l = -m_n r^0 - m_e^2 + \frac{1}{2} \Delta
m^2$, $k r = -m_n l^0 + \kappa^2 + \frac{1}{2} \Delta m^2$, $p r = m_n
r^0$, $l p = m_n l^0$, $l r = m_n (l^0 + r^0) - \frac{1}{2} \Delta
m^2$ and $\Delta m^2=m_{n}^{2}-m_{p}^{2}+m_{e}^{2}-\kappa^2$.
\begin{figure}
  \begin{center}
  \psfig{width=.5\textwidth,figure=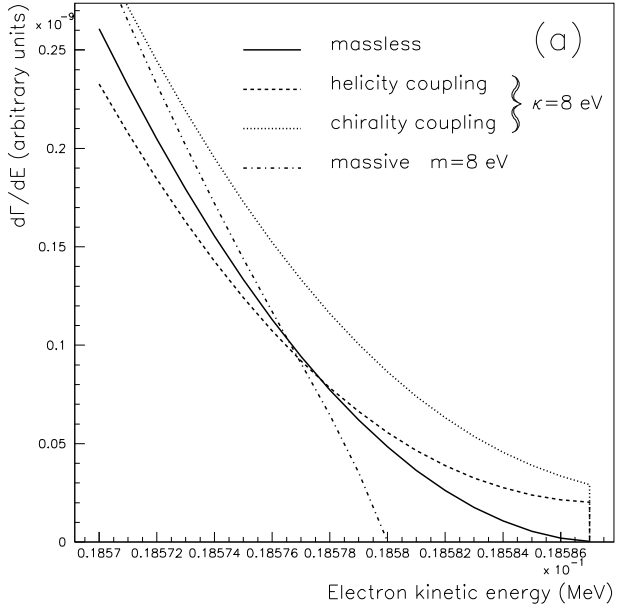}
  \psfig{width=.5\textwidth,figure=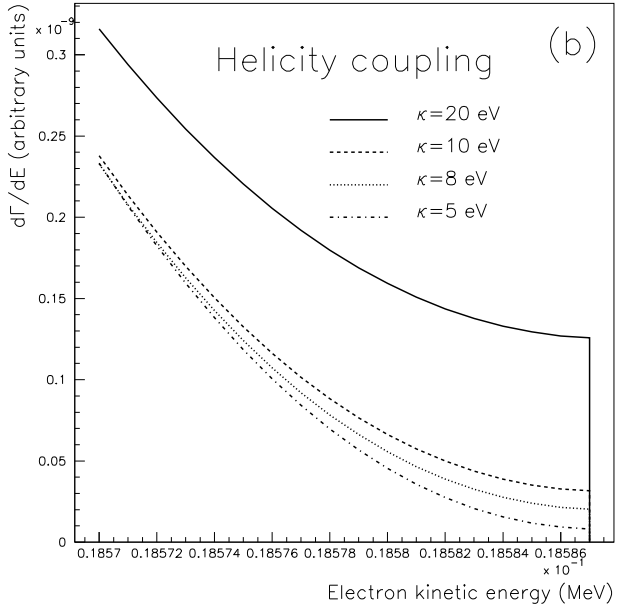}
  \end{center}
  \caption{(a) Differential electron energy spectra in the vicinity
    of the end point, for tritium decay with: a tachyonic antineutrino
    of mass $\kappa=8\;\eV$, massless neutrino and massive neutrino of
    mass $m=8\;\eV$; (b) as above for a tachyonic electron
    antineutrino with the helicity coupling, for a range of tachyonic
    masses, $\kappa$.}
  \label{ryc:1stpicture}
\end{figure}

\subsection{Electron energy spectrum}

We have calculated differential electron energy spectra, $d\Gamma
/dl^{0}$, in tritium decay using the following values for the masses
of $^{3}\mathrm{H}$ and $^{3}\mathrm{He}$: $m_{n} = 2809.94\;\MeV$ and
$m_{p} = 2809.41\;\MeV$, respectively (hereafter we use $l^{0}\equiv
E$).  The corresponding value for the electron end point kinetic
energy in the preferred frame is $T_{\rm max} = 18587.56\;\eV$.
Differential electron energy spectra, $d\Gamma /d E$, corresponding to
decays with a massive, massless and tachyonic neutrino, in the
vicinity of the endpoint, are shown in Fig.~\ref{ryc:1stpicture}a.
The tachyonic spectra for both couplings near the end point rise above
that for the massless neutrino.  Moreover, they terminate at $T =
T_{\rm max}$ with a quasi step: the function $d\Gamma/d E$ decreases
linearly to zero over the energy interval of $2\kappa p_{\rm
  max}/m_{n}$ (where $p_{\rm max}$ denotes the maximal electron
momentum) which in the tritium decay amounts to $\approx 10^{-3}\;\eV$
for $\kappa=8\;\eV$.  The magnitude (height) of the step depends on
the choice of coupling as well as on the value of $\kappa$, as can be
seen in Fig.~\ref{ryc:1stpicture}a,b.
\begin{figure}
  \begin{center}
    \psfig{width=.5\textwidth,figure=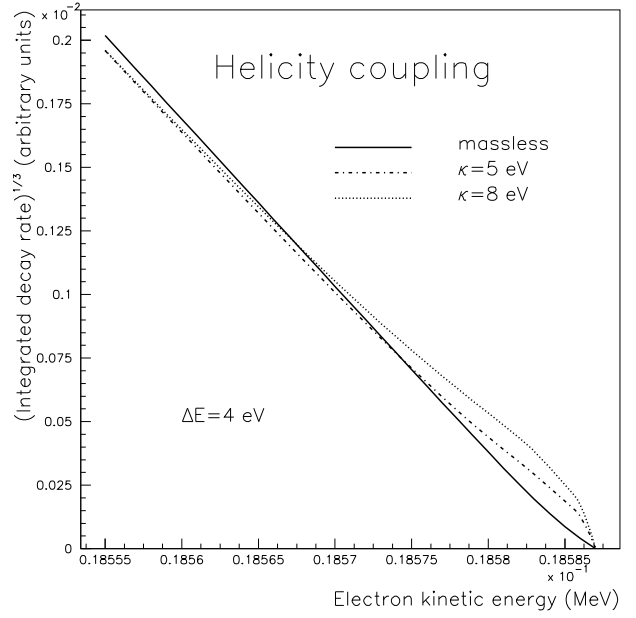}
  \end{center}
  \caption{Linearised (cube root)  integral electron energy spectrum
    with folded experimental resolution function (see text).}
  \label{ryc:FOUR}
\end{figure}

Thus if the neutrino were a tachyon, one would expect an excess of the
counting rate near the end point, i.e. an effect qualitatively similar
to the one actually observed.  Since the detectors used in Mainz and
Troitsk are integrating spectrometers, we integrated the electron
energy spectrum given by (\ref{eq:dGdE1}), folding in a simplified
experimental resolution function used in these
experiments~\cite{titov97}, with energy resolution $\Delta E=4\;\eV$
(not accounting for the final state energy spectrum).  The resulting
linearised (cube root) electron energy spectrum near the endpoint is
shown in Fig.~\ref{ryc:FOUR}.  There is a striking similarity of the
predicted shape and that observed in the Troitsk data~\cite{troitsk}d
(which is the only published linearised spectrum).  We also verified
that the {\it end point effect\/} of the observed magnitude could have
hardly been discovered in earlier measurements which had much poorer
energy resolution.

For practical purposes the rigorous but complicated expressions for
the electron energy spectra~(\ref{eq:dGdE1}) may be approximated in
order to write them in terms of the variable $(E_{\rm max}-E)$ and the
electron momentum $p$. The simplified form, valid under the
condition $E\leq E_{\rm max}$, for the helicity coupling reads:
\begin{multline}
  \frac{d\Gamma}{d E}= \frac{G_{F}}{2 \pi^{3}} \Bigl[ \kappa m_e (1-3
    g_{A}^{2}) \\ + (1 + 3 g_{A}^{2}) E \sqrt{(E_{\rm
        max}-E)^2 + \kappa^{2}} \Bigr] \times \\ \times p
  \sqrt{(E_{\rm max}-E)^{2} + \kappa^{2}}
\end{multline}
and for the chirality coupling:
\begin{multline}
  \frac{d\Gamma}{d E}= \frac{G_F}{4 \pi^{3}} (1 + 3 g_{A}^{2}) E p
  \Bigl[(E_{\rm max}-E)^{2} \\
  + (E_{\rm max}-E) \sqrt{ (E_{\rm max}-E)^{2} + \kappa^{2}} +
    \kappa^{2} \Bigr],
\end{multline}
with the additional  explicite condition that 
$\frac{d\Gamma}{d E} = 0$ for  $E>E_{\rm max}$ since
the approximated spectra do not vanish  at  $E=E_{\rm max}$ (step).

\section{Preferred frame and time dependent effects}\label{sec:Time_dep}

An interesting property of amplitudes for processes involving
tachyonic neutrinos, in particular for the beta decay, is their
dependence on the velocity four-vector of the preferred frame, $u$
(Eq.~(\ref{eq:beta4})).  On the grounds of cosmological considerations
one might expect that a frame in which the cosmic microwave background
radiation (CMBR) is isotropic is a natural candidate for the preferred
frame.  In such a case results derived in previous sections are
sufficiently precise because the Solar System is almost at rest
relatively to the CMBR\footnote{Velocity deduced from the dipole
  anisotropy in temperature amounts to about $350\;{\rm km}/{\rm s}$
  \protect{\cite{lubin83}}.}.

Consider a certain configuration of the final state particles momenta
in a beta decay which occurs in a reference frame moving with velocity
$\vec{u}$ with respect to the preferred frame.  The maximal kinetic
energy of the electron, $T_{\rm max}$, depends on $\beta=|\vec{u}|/c$ and
$\cos\omega$, where $\omega$ is the angle between the neutrino
momentum and the vector $\vec{u}$:
\begin{equation}\label{eq:Tmax1}
  T_{\rm max}(\beta,\cos\omega) = T_{\rm max} - \Delta T_{\rm
    max}(\beta,\cos\omega)
\end{equation}
where
\begin{equation}\label{eq:Tmax2}
  \Delta T_{\rm max}(\beta,\cos\omega) = \frac{\kappa\beta
    \cos\omega}{\sqrt{1-\beta^{2}\cos^{2}\omega}}.
\end{equation}
Momenta of the final state particles in the $\beta$ decay are aligned
at the end point and thus the angle $\omega$ may be expressed by the
angle corresponding to the electron.  Assume for simplicity that the
electron spectrum is measured in an ideal spectrometer in which
electrons are moving along the spectrometer axis. Thus the angle
$\omega$ between this axis and the vector $\vec{u}$ changes with time
due to Earth's rotation and day -- night variations of $T_{\rm max}$ are
expected.  If we identify the preferred frame with the CMBR
($\beta\approx 10^{-3}$) we obtain $\Delta T_{\rm max} < 10^{-2}\;\eV$
for the tachyonic electron neutrino mass of a few $\eV$, i.e.  an
effect undetectable at present.  If however the velocity of the
preferred frame were large ($\beta>0.1$), the expected variation of
the end point energy would be of order $\eV$.

\section{Summary and conclusions}

We have shown that the electron energy spectrum in a beta decay with a
tachyonic neutrino rises above that for the massless neutrino and ends
with a quasi step at $E=E_{\rm max}$.  This feature may explain the
excess of counts observed in tritium decay in the vicinity of the end
point if the neutrino were a tachyon.  Our prediction and the
measurement of Troitsk show a remarkable similarity when presented on
a linearised plot.  Since the neutrino field is an eigenvector of the
helicity operator with the eigenvalue $\frac{1}{2}$ -- {\em helicity
  coupling offers a natural explanation of the V--A structure of the
  weak leptonic current in neutrino interactions}.  Certain
preliminary considerations concerning the hypothesis of tachyonic
neutrinos may be found elsewhere~\cite{our}.

\section*{Acknowledgement}
  We wish to thank K.A. Smoli\'nski and P. Caban for their
  contribution to numerical calculations; B. Jeziorski for discussions
  concerning molecular final states; E.~Otten, J.~Bonn and
  Ch.~Weinheimer for numerous discussions, information about the
  detector and the results of the Mainz experiment as well as comments
  concerning the manuscript; V.M.~Lobashev and N.A.~Titov for
  information concerning the Troitsk experiment.


\begin{thebibliography}{99}
\bibitem{mainz} Mainz ${^3}{\rm T}$ decay experiment:\\
  a. Ch. Weinheimer, in {\it Proc.\ XVIII Conf.\ on Neutrino
    Physics and Astrophysics (NEUTRINO98), Takayama, Japan, 1998},
  (to appear)\\
  b. J. Bonn, in {\it Proc.\ Int.\ Conf.\ on High Energy Physics
    (HEP97), Jerusalem, 1997}\\
  c. H. Barth et al., in {\it Proc.\ of the XVI Workshop on Weak
    Interactions and Neutrinos (WIN97), Capri, Italy, 1997}, Nucl.\ 
  Phys.\ B (Proc.\ Suppl.) {\bf 66}, 183 (1998)\\
  d. J. Bonn, in {\it Proc.\ XVII Conf.\ on Neutrino Physics and
    Astrophysics (NEUTRINO96), Helsinki, Finland, 1996}, p.~259\\
  e. Ch. Weinheimer et al., Phys.\ Lett.\ B {\bf 300}, 210 (1993)
\bibitem{troitsk} Troitsk ${^3}{\rm T}$ experiment:\\
  a. V.M. Lobashev, in {\it Proc.\ XVIII Conf.\ on Neutrino
    Physics and Astrophysics (NEUTRINO98), Takayama, Japan, 1998},
  (to appear)\\
  b. V.M. Lobashev, in {\it Proc.\ of the XVI Workshop on Weak
    Interactions and Neutrinos (WIN97), Capri, Italy, 1997}, Nucl.\ 
  Phys.\ B (Proc.\ Suppl.) {\bf 66}, 187 (1998)\\
  c. V.M. Lobashev, in {\it Proc.\ XVII Conf.\ on Neutrino
    Physics and Astrophysics (NEUTRINO96), Helsinki, Finland, 1996},
  p.~264\\
  d. A.I. Belesev et al., Phys.\ Lett.\ B {\bf 350}, 263 (1995)
\bibitem{stoeffl95} W. Stoeffl, D. Decman et al., Phys.\ Rev.\ Lett.
  {\bf 75}, 3237 (1995) \bibitem{other} H. C. Sun et al., Chin.\ J.
  Nucl.\ Phys. {\bf 15}, 261
  (1993)\\
  E. Holzschuh et al., Phys.\ Lett.\ B {\bf 287}, 381 (1992)\\
  H. Kawakami et al., Phys.\ Lett.\ B {\bf 256}, 105 (1991)\\
  R.G.H. Robertson et al., Phys.\ Rev.\ Lett.\ {\bf 67}, 957 (1993)
\bibitem{fsi} P. Froelich et al., Phys.\ Rev.\  Lett.\ {\bf 71}, 2871 (1993)\\
  H.C. Sun et al., Int.\ J. Mod.\ Phys.\ A {\bf 10}, 2841 (1995)\\
  Workshop: `The Tritium $\beta$ Decay Spectrum: The negative
  $m^{2}_{\nu_{e}}$ Issue', Harvard--Smithsonian Center for
  Astrophysics, Cambridge, Massachusetts, USA, 1996\\
  S. Jonsell, H.J. Monkhorst, Phys.\ Rev.\ Lett.\ {\bf 76}, 4476 (1996)\\
  B. Jeziorski (private communications)
\bibitem{stat} L.A. Khalfin, PDMI-8/1996 (unpublished)\\
  M. Roos, L.A. Khalfin, HU-TFT-96-13 {\tt hep-ex/9605008}
  (unpublished)
\bibitem{unconv} R.N. Mohapatra, S. Nussinov, Phys.\ Lett.\ B {\bf
  398}, 63 (1997)\\ 
  J.I. Collar, {\tt hep-ph/9611420} (unpublished)
\bibitem{jaremb} J. Rembieli\'nski, Phys.\ Lett.\ A {\bf 78}, 33 (1980)\\
  J. Rembieli\'nski, KFT U\L\ 2/94 {\tt hep-th/9410079} (unpublished)\\
  J. Rembieli\'nski, KFT U\L\ 5/94 {\tt hep-th/9411230} (unpublished)\\
  J. Rembieli\'nski, Int.\ J. Mod.\ Phys.\ A {\bf 12}, 1677 (1997)
\bibitem{ct} T. Chang, Phys.\ Lett.\ A  {\bf 70}, 1 (1979)\\
  T. Chang, J. Phys.\ A {\bf 12}, L203 (1979)\\
  T. Chang, J. Phys.\ A  {\bf 13}, L207 (1980)\\
  F.R. Tangherlini,  N. Cim.\ Suppl.\ {\bf 20}, 1 (1961)
\bibitem{chodos85} A. Chodos, A.I. Hauser, V.A. Kostelecky, Phys.\
  Lett.\ B {\bf 150}, 431 (1985)
\bibitem{titov97} Ch. Weinheimer, N.A. Titov (private communications)
\bibitem{lubin83} P.M. Lubin et al., Phys.\ Rev.\ Lett.\ {\bf 50},
  616 (1983)\\
 D.J. Fixsen et al., Phys.\ Rev.\ Lett.\ {\bf 50}, 620 (1983)
\bibitem{our} J. Ciborowski, J. Rembieli\'nski, in {\it Proc.\ XXVIII Int.\
  Conf.\ on High Energy Physics (ICHEP96), Warsaw, 1996},
  edited by  Z. Ajduk, A.K. Wr\'oblewski (World Scientific,
  Singapore 1997), p.~1247\\
  J. Ciborowski, J. Rembieli\'nski, {\tt hep-ph/9607477} (unpublished)\\
  J. Ciborowski, J. Rembieli\'nski, in {\it Proc.\ Int.\ Conf.\
  on High Energy Physics (HEP97), Jerusalem, 1997} (to appear) 
\end{thebibliography}
\end{document}